\documentclass[aps, 10pt, prl,notitlepage, twocolumn, superscriptaddress,nofootinbib]{revtex4-2}

\usepackage{bm}
\usepackage{amsfonts}
\usepackage{latexsym}
\usepackage[latin1]{inputenc}
\usepackage{graphicx}
\usepackage{amsmath}
\usepackage{palatino}
\usepackage{mathpazo}
\usepackage{booktabs}
\usepackage{dcolumn}

\def\jnl@style{\it}
\def\aaref@jnl#1{{\jnl@style#1}}

\def\aaref@jnl#1{{\jnl@style#1}}

\def\aj{\aaref@jnl{AJ}}                   
\def\apj{\aaref@jnl{ApJ}}                 
\def\apjl{\aaref@jnl{ApJ}}                
\def\apjs{\aaref@jnl{ApJS}}               
\def\apss{\aaref@jnl{Ap\&SS}}             
\def\aap{\aaref@jnl{A\&A}}                
\def\aapr{\aaref@jnl{A\&A~Rev.}}          
\def\aaps{\aaref@jnl{A\&AS}}              
\def\mnras{\aaref@jnl{Mon.~Not.~Roy.~Astron.~Soc.}}             
\def\prd{\aaref@jnl{Phys.~Rev.~D}}        
\def\prc{\aaref@jnl{Phys.~Rev.~C}}  
\def\prl{\aaref@jnl{Phys.~Rev.~Lett.}}    
\def\qjras{\aaref@jnl{QJRAS}}             
\def\skytel{\aaref@jnl{S\&T}}             
\def\ssr{\aaref@jnl{Space~Sci.~Rev.}}     
\def\zap{\aaref@jnl{ZAp}}                 
\def\nat{\aaref@jnl{Nature}}              
\def\aplett{\aaref@jnl{Astrophys.~Lett.}} 
\def\apspr{\aaref@jnl{Astrophys.~Space~Phys.~Res.}} 
\def\physrep{\aaref@jnl{Phys.~Rep.}}      
\def\physscr{\aaref@jnl{Phys.~Scr}}       
\def\commat{\aaref@jnl{Comm.~Math.~Phys.}}              
\def\science{\aaref@jnl{Science}}               
\def\cqg{\aaref@jnl{Classical Quant.~Grav.}}            
\def\jpcs{\aaref@jnl{JPCS}}                                     
\def\ijmpd{\aaref@jnl{Int.~J.~Mod.~Phys.~D}}                    
\def\grg{\aaref@jnl{Gen.~Relat.~Gravit.}}               
\def\rpp{\aaref@jnl{Rep.~Prog.~Phys.}}          
\def\npa{\aaref@jnl{Nucl.~Phys.~A}}        
\def\lrr{\aaref@jnl{Living Rev.~Rel.}}                   
\def\jcap{\aaref@jnl{J.~Cosmology Astropart.~Phys.}}    
\def\rmp{\aaref@jnl{Rev.~Mod.~Phys.}}   

\allowdisplaybreaks[1]

\begin{document}

	\title{Beyond the spontaneous scalarization: New fully nonlinear dynamical mechanism for formation of scalarized black holes}
	
	\author{Daniela D. Doneva}
	\email{daniela.doneva@uni-tuebingen.de}
	\affiliation{Theoretical Astrophysics, Eberhard Karls University of T\"ubingen, T\"ubingen 72076, Germany}
	\affiliation{INRNE - Bulgarian Academy of Sciences, 1784  Sofia, Bulgaria}

	\author{Stoytcho S. Yazadjiev}
	\email{yazad@phys.uni-sofia.bg}
	\affiliation{Theoretical Astrophysics, Eberhard Karls University of T\"ubingen, T\"ubingen 72076, Germany}
	\affiliation{Department of Theoretical Physics, Faculty of Physics, Sofia University, Sofia 1164, Bulgaria}
	\affiliation{Institute of Mathematics and Informatics, 	Bulgarian Academy of Sciences, 	Acad. G. Bonchev St. 8, Sofia 1113, Bulgaria}


\begin{abstract}
In the present letter we show the existence of a fully nonlinear dynamical mechanism for the formation of scalarized black holes  which is different from the spontaneous scalarization. We consider  a class of scalar-Gauss-Bonnet gravity theories  within which no tachyonic instability can occur. Although the Schwarzschild black holes are  linearly stable against scalar perturbations, we show dynamically that for certain choices of the coupling function they are unstable against nonlinear scalar perturbations.  This nonlinear instability leads to the formation of new black holes with scalar hair. The fully nonlinear and self-consistent  study of the equilibrium black holes reveals that the spectrum of solutions is more complicated and more than one scalarized branch can exist. We have also considered classes of scalar-Gauss-Bonnet theories where both the standard and the nonlinear scalarization can be present, and they are smoothly connected that completes in an interesting way the picture of black hole scalarization.  The fully nonlinear (de)scalarization of a Schwarzschild black hole will always happen with a jump because the stable ``scalarized branch'' is not continuously connected to the Schwarzschild one that can leave clear observational signatures.  
\end{abstract}
	
	\maketitle
	
\textit{Introduction.} Spontaneous scalarization is a dynamical mechanism for endowing black holes (and other compact objects) with scalar hair without altering the predictions in the weak field limit \cite{Damour1993,Damour1996,Doneva:2017bvd,Silva:2017uqg,Antoniou:2017acq}. It is a strong gravity phase transition triggered by tachyonic instability  due to the non-minimal coupling between the scalar field(s) and the spacetime curvature (or matter). The realistic black hole spontaneous scalarization was mainly studied within the  scalar-Gauss-Bonnet (sGB) gravity defined by the action  	
\begin{eqnarray}\label{GBA}
S=&&\frac{1}{16\pi}\int d^4x \sqrt{-g} 
\Big[R - 2\nabla_\mu \varphi \nabla^\mu \varphi 
+ \lambda^2 f(\varphi){\cal R}^2_{GB} \Big] .\label{eq:quadratic}
\end{eqnarray}
where $R$ is the Ricci scalar with respect to the spacetime metric $g_{\mu\nu}$, $\varphi$ denotes the scalar field  with a coupling function  $f(\varphi)$, $\lambda$ is the so-called Gauss-Bonnet coupling constant having  dimension of $length$ and ${\cal R}^2_{GB}=R^2 - 4 R_{\mu\nu} R^{\mu\nu} + R_{\mu\nu\alpha\beta}R^{\mu\nu\alpha\beta}$ is the Gauss-Bonnet invariant with $R_{\mu\nu\alpha\beta}$ being the 
curvature tensor. Black hole spontaneous scalarization was extensively studied (see e.g. \cite{Doneva:2017bvd,Silva:2017uqg,Antoniou:2017acq,Minamitsuji:2018xde,Silva:2018qhn,Doneva:2019vuh,Macedo:2019sem,Blazquez-Salcedo:2021npn,Antoniou:2021zoy}), including the rotating case \cite{Cunha:2019dwb,Collodel:2019kkx} and the spin-induced scalarization \cite{Dima:2020yac,Doneva:2020nbb,Doneva:2020kfv,Herdeiro:2020wei,Berti:2020kgk}. It turns out that these black holes are energetically favorable over the GR solutions and they are stable \cite{Blazquez-Salcedo:2018jnn,Blazquez-Salcedo:2020rhf,Blazquez-Salcedo:2020caw}.  The nonlinear dynamics of scalarized black holes in sGB gravity, including mergers and stellar core-collapse, was examined in \cite{Ripley:2020vpk,Silva:2020omi,Doneva:2021dqn,Kuan:2021lol,East:2021bqk}. Spontaneous scalarization was also considered in other alternative theories of gravity \cite{Herdeiro:2018wub,Andreou:2019ikc,Gao:2018acg,Doneva:2021dcc,Zhang:2021btn,Myung:2021ztl}.

In the present work we would like  to go beyond the spontaneous scalarization. We pose the following problem. Let us consider  sGB theories
with coupling functions  that do not allow for a tachyonic instability to occurs, i.e. sGB theories which do not exhibit spontaneous scalarization. Then we ask: Is there another dynamical mechanism for producing scalar hair of the black holes? The purpose of this letter is 
to answer this question. 

We identify  a class of sGB theories with coupling between the scalar field and the Gauss-Bonnet invariant
that {\it can not exhibit} spontaneous scalarization for  black holes but admits all of the stationary solutions of general relativity,  
including the Schwarzschild black holes. Within the mentioned class of sGB theories the Schwarzschild black holes are  linearly stable.  However, we show dynamically that they are unstable against nonlinear scalar perturbations within the sGB gravity  and that the nonlinear instability leads to the formation of new black holes with scalar hair. 
We also go one step further and consider sGB theories which allow the simultaneous existence of both standard and nonlinear scalarization and show the smooth connection between the two. This completes the picture of black hole scalarization in sGB gravity opening possibilities for interesting astrophysical manifestations.

 \textit{Nonlinear instability of the Schwarzschild black hole and dynamical formation of scalarzied black holes.}
We consider sGB theories which can not exhibit tachyonic instability. Therefore, we impose the following conditions on the Gauss-Bonnet coupling function
\begin{eqnarray}
f(0)=0, \; \; \frac{df}{d\varphi}(0)=0, \;\;    \frac{d^2f}{d\varphi^2}(0)=0. \label{eq:CouplingFuncConditions}
\end{eqnarray}
The first condition can always be imposed since the field equations include only the first derivative of the coupling function but not the coupling function itself. The second condition guarantees the Schwarzschild solution is also a black hole solution to the equations of the  sGB gravity with $\varphi=0$. It is not difficult one to see that the third condition  $\frac{d^2f}{d\varphi^2}(0)=0$ imposed on the Gauss-Bonnet coupling function leads to the fact that no tachyonic instability is possible -- the Schwarzschild solution is stable against linear scalar perturbations. Indeed, in the case of  $ \frac{d^2f}{d\varphi^2}(0)=0$ the equation governing the scalar perturbations is just the scalar wave equation and it is a well-known fact the Schwarzschild black hole is stable against the linear perturbations of a massless scalar perturbations. Therefore, if black holes with scalar hair exist, they should form  a new black hole phase coexisting with the usual Schwarzschild black hole phase.  It will be called a scalarized black hole phase.     
	 
We will investigate how the scalarized phases of the Schwarzschild black hole can be dynamically formed and whether they are stable. We shall base our study on an approximate model which is free from heavy technical complications but preserves the leading role of the nonlinearity associated with the coupling function. More precisely, we adopt the decoupling limit approximation where the spacetime geometry is kept fixed and the whole dynamics is governed by the nonlinear equation for the scalar field. This dynamical model is a very good approximation when the energy of the scalar hair is less than the full mass of the black hole \cite{Doneva:2021dqn} and we will follow the methodology described in \cite{Doneva:2020nbb,Doneva:2021dqn}.

The conditions \eqref{eq:CouplingFuncConditions} are easily satisfied by a coupling function containing power of $\varphi$ higher than two. In the present letter we are focusing on $Z_2$ symmetric theories and the first two possible choices are $\varphi^4$ and $\varphi^6$. In the case when the coupling functions are directly proportional to $\varphi^4$ and $\varphi^6$, though, we could not find any stable solutions similar to the case of standard scalarization in sGB gravity \cite{Minamitsuji:2018xde,Silva:2018qhn,Doneva:2019vuh,Macedo:2019sem}. Perhaps the most straightforward and easy to handle from a numeral point of view is to consider an exponential coupling function of the form:
\begin{eqnarray}\label{eq:coupling_phi4}
&& f_1(\varphi)= \frac{1}{4\kappa}\left(1- e^{-\kappa\varphi^4}\right), \;\;\;   
f_2(\varphi)= \frac{1}{6\kappa}\left(1- e^{-\kappa\varphi^6}\right) 
\end{eqnarray}   
where $\kappa$ is a parameter. In the main text we present our results for the coupling functions $f_1(\varphi)$. The results for $f_2(\varphi)$ are qualitatively similar and they are given in the Supplemental material for completeness.

\begin{figure}
	\includegraphics[width=0.45\textwidth]{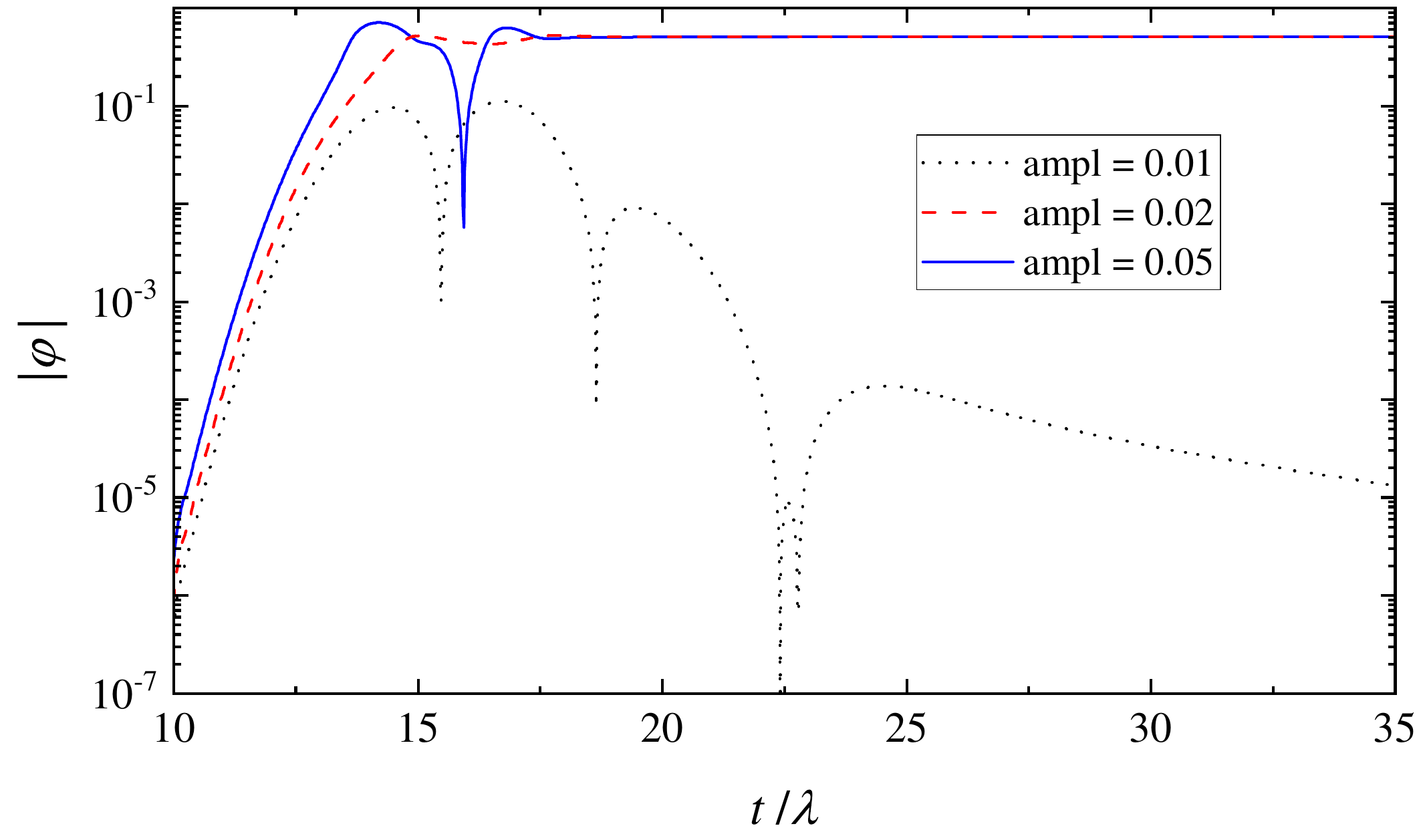}
	\caption{The time evolution of the scalar field on the background of a Schwarzschild black hole with mass $M/\lambda=0.1$ and $\kappa=50$. The initial data is a Gauss pulse with amplitudes 0.001, 0.002 and 0.005, dispersion $\sigma/\lambda=1$, located at coordinate distance $x/\lambda=12$. The extraction of the signal is at  $x/\lambda=10$. The coupling function is $f_1(\varphi)$. }
	\label{fig:phi_t_phi4_b50}
\end{figure}

For these choices of the coupling function, the Schwarzschild black hole is stable only with respect to linear perturbations and it might be unstable with respect to larger nonlinear perturbations. By evolving in time the nonlinear scalar field equation  on the Schwarzschild background, it turns out that there is a threshold amplitude of the perturbation above which the Schwarzschild black hole loses stability and a scalarized phase forms. This is demonstrated in Fig. \ref{fig:phi_t_phi4_b50} where the time evolution of the  scalar field on the Schwarzschild background is shown for several amplitudes of the initial perturbation. As one can see, for small amplitudes the standard quasinormal modes and exponential decay is observed. \textbf{Once the amplitude exceed a certain threshold, the scalar field first grows due to the energy carried out by the initial perturbation and afterward it stabilizes to a constant that is the formation of a scalarized phase.} Through this approach only the formation of a stable scalarized phase can be followed and independent on the perturbation, we observed the formation of only one probably stable scalarized phase. Clearly, other scalarized phase can exist that can be constructed by solving the full system of static field equations.

\begin{figure}
	\includegraphics[width=0.45\textwidth]{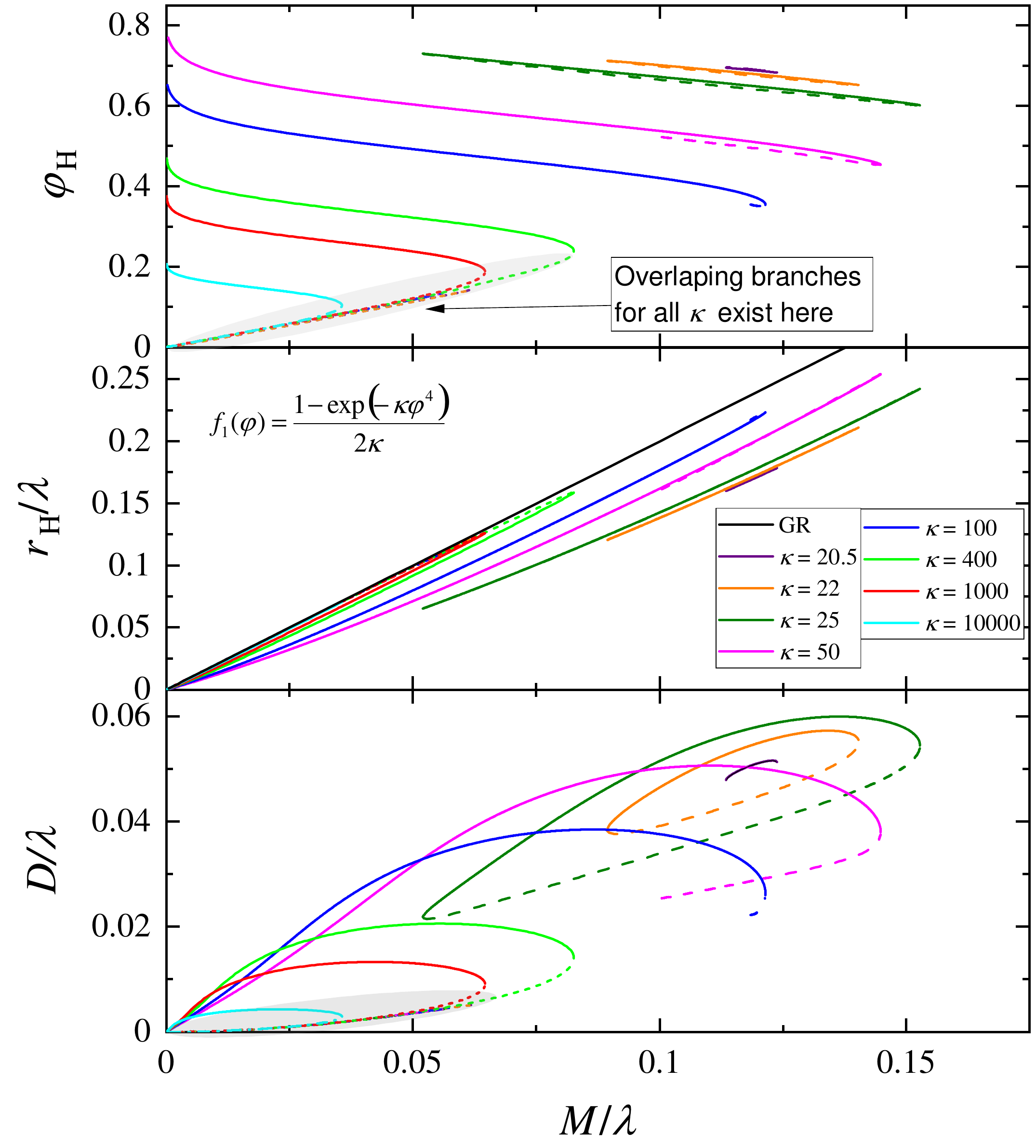}
	\caption{The scalar field at the horizon (top panel), the mass (middle panel) and the scalar charge (bottom panel) as functions of the black hole horizon mass for the $f_1(\varphi)$ coupling. }
	\label{fig:M_phiH_D_phi4}
\end{figure}

\textit{Nonlinear scalarized phases.}
After gaining insight where stable scalarized phases can exist and how they form dynamically, the next step is to study the full spectrum of both stable and unstable solution. For this purpose we solve the fully nonlinear and self-consistent system of reduced static field equations \cite{Doneva:2017bvd}, that are commented in detail in the Supplemental material. The scalarized phases of the Schwarzschild black hole are displayed in Fig. \ref{fig:M_phiH_D_phi4} where the scalar field on the horizon, the mass and the scalar charge are given for  several values of $\kappa$. Here the scalar charge $D$ is defined as the coefficient in the leading order asymptotic of the scalar field at infinity	$\varphi |_{r \rightarrow \infty} \sim D/r$.
All dimensional quantities are normalized with respect to $\lambda$. Only the solutions with $\varphi>0$ are displayed but one should keep in mind that the field equations, with this choice of the coupling function \eqref{eq:coupling_phi4}, are symmetric with respect to a change of the sign of $\varphi$ and thus solutions with the same metric functions but opposite sign of $\varphi$ also exist. 

The qualitative picture changes for different $\kappa$ and there are three possible cases. The first one realizes for large enough $\kappa$ (e.g. $\kappa=400$) when two branches of solutions exist and both of them start from the $M=0$ limit -- one with vanishing (dotted line) and one with increasing (solid line) scalar field at the horizon for $M \rightarrow 0$. The nonlinear evolution of the scalar field suggests that the latter branch is the stable one.

For intermediate $\kappa$ (e.g. $\kappa=100$) up to three branches can exist -- the lower one (dotted line) having $\varphi_{H} \rightarrow 0$ when $M\rightarrow 0$ is terminated at some finite $M/\lambda$ where the solutions disappear due to violation of the scalar field regularity condition at the horizon \cite{Doneva:2017bvd} (see the Supplemental material). The upper branch (solid line) starts from zero mass and with the increase of $M/\lambda$ the scalar field on the horizon decreases until a maximum of $M/\lambda$ is reached. At that point the branch merges with a third middle branch of solutions (dashed line). The uppermost branch is the only stable one.

For small $\kappa$ (e.g. $\kappa=50$) there are also three branches of solutions but the upper two merge at two points -- at their beginning and at the end of the branch. Again, the stable branch is the one depicted with solid line. 

This picture is clearly different from the Einstein-Maxwell-scalar gravity \cite{Blazquez-Salcedo:2020nhs} where up to two scalarized branches exist and one of them originates from the extremal Reissner-Nordstr\"om solution. Since here no extremal solution exists,  most probably the scalarized phases are connected with a solitonic-like solutions in sGB gravity of the type considered in \cite{Kleihaus:2019rbg,Kleihaus:2020qwo}. 

As one can see in Fig. \ref{fig:M_phiH_D_phi4}, there is a clustering of the lower mass black hole branches for different $\kappa$ in the gray shaded area. With the decrease of $\kappa$ the potentially stable branch, depicted with solid line,  gets shorter until it completely disappears at roughly $\kappa \approx 20$. On the contrary, scalarized branches exist for arbitrary large $\kappa$, but they span a decreasingly small range of masses.

The radius of the horizon can differs significantly from the Schwarzschild one especially for small $\kappa$ and gets closer to GR with the increase of $\kappa$. This behavior is similar to the standard scalarization \cite{Doneva:2018rou} where also the differences from the Schwarzschild black hole are reduced for larger $\kappa$. 
As one can see in the top and middle panels of Fig. \ref{fig:M_phiH_D_phi4} the differences especially between the two upper branches (for a fixed $\kappa$) are very small and seem negligible. If one looks at the scalar charge in the lower panel, though, a clear difference between the branches is observed.

Let us comment further on the stability of the solutions. As commented, the Schwarzschild solution is always linearly stable for the coupling \eqref{eq:coupling_phi4}. According to the scalar field evolution only the branch depicted with solid lines in Fig.  \ref{fig:M_phiH_D_phi4} is stable. Important information can be also inferred by studying the black hole thermodynamical properties. The entropy at the horizon can be defined as $S_H =  A_H/4 + 4\pi \lambda^2 f(\varphi_{H})$ \cite{Doneva:2017bvd}. Among the scalarized phases (for fixed $\kappa$), the one depicted with solid line in Fig. \ref{fig:M_phiH_D_phi4} has the largest entropy and hence it is most probably the only stable one. For small enough $\kappa$ the whole scalarized phase branch has larger entropy compared to the Schwarzschild phase and is thus thermodynamically preferred. For larger $\kappa$ and around the maximum $M/\lambda$, though, a small part of the scalarized branch can have lower entropy compared to the Schwarzschild one.

\begin{figure}
	\includegraphics[width=0.45\textwidth]{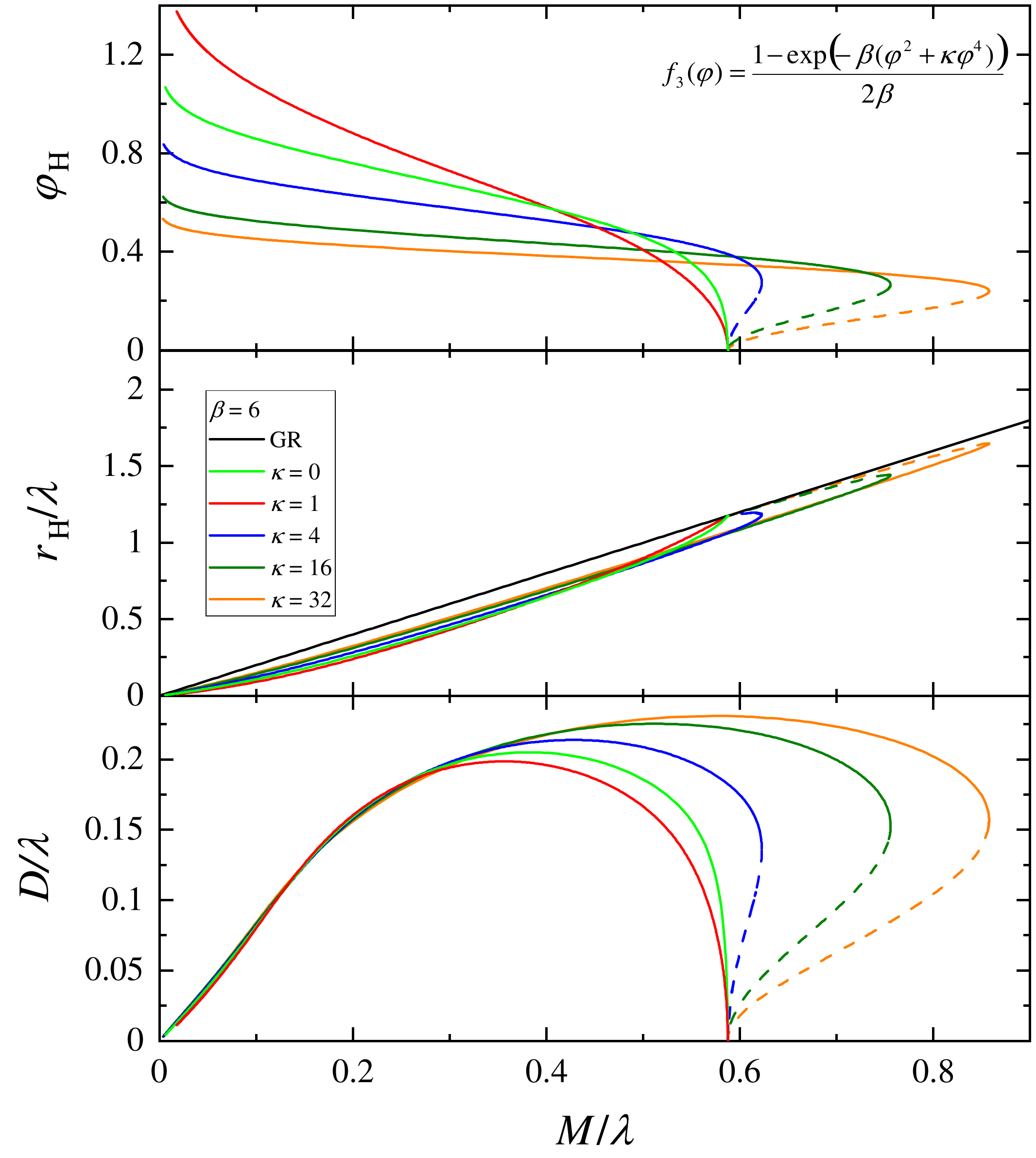}
	\caption{The scalar field at the horizon (top panel), the mass (middle panel) and the scalar charge (bottom panel) as functions of the black hole horizon mass for the $f_3(\varphi)$ coupling. The unstable branches are depicted with dotted lines}
	\label{fig:M_phiH_D_phi2_phi4}
\end{figure}

\textit{Connecting nonlinear scalarized phases and standard scalarization}
The coupling functions \eqref{eq:coupling_phi4} are chosen as the simplest ones allowing for the existence of nonlinear scalarized phases. Here we will study a more complicated coupling allowing for the co-existence of standard and fully nonlinear scalarization at the same time:
\begin{eqnarray}\label{eq:coupling_phi2_phi4}
f_3(\varphi)= \frac{1}{2\beta}\left(1- e^{-\beta(\varphi^2 + \kappa \varphi^4)}\right).  \nonumber
\end{eqnarray}   
As one can easily check, $f_3(\varphi)$ does not satisfy anymore all the conditions \eqref{eq:CouplingFuncConditions} and the Schwarzschild solution is linearly unstable below certain mass of the black hole. The branches of solutions for $\beta=6$ and varying $\kappa$ are depicted in Fig. \ref{fig:M_phiH_D_phi2_phi4}. For small enough $\kappa$ (in the figure $\kappa=0$ and $\kappa=1$) only one branch of scalarized solutions exists that resembles the standard one reported in \cite{Doneva:2017bvd}. With the increase of $\kappa$ this branch changes slope (it turns right instead of left) that is normally a sign of instability \cite{Macedo:2019sem}. At the point of termination of this ``standard'' unstable scalarized branch (depicted with dotted line), though, a new one appears (depicted with solid line) that has properties resembling closely the fully nonlinear scalarization observed in the previous sections. Namely, for black hole masses larger than the bifurcation point where the Schwarzschild solution is linearly stable, the scalar hair can be excited only if sufficiently high scalar field perturbation is imposed. The time evolution of the scalar field for the coupling \eqref{eq:coupling_phi2_phi4} is qualitatively the same as the one shown in Fig. \ref{fig:phi_t_phi4_b50} for the coupling \eqref{eq:coupling_phi4} and that is why we will not comment on it separately.

This shows a smooth connection between the ``standard'' and fully nonlinear scalarization that completes the picture. Moreover, the scalarized phase branch (depicted with solid line in Fig. \ref{eq:coupling_phi2_phi4}) has higher entropy compared to the dashed line branch and also to the Schwarzschild one for most of the parameter range. This also speaks in favor of its stability.

\textit{Discussion}
In the present letter we have considered the possibilities to go beyond the standard scalarization of black holes in scalar-Gauss-Bonnet gravity.
 We showed the existence of a fully nonlinear dynamical mechanism for the formation of scalarized black holes which is different from the spontaneous scalarization. We considered types of coupling functions for which the Schwarzschild black hole is still a linearly stable solution of the field equations but for certain ranges of the parameters scalarized phases of the Schwarzschild black hole can exist. The reason for the appearance of such phases is that even though the Schwarzschild phase is stable against small (linear) perturbations, this stability can be lost for larger amplitudes of the perturbations that will bring us in the nonlinear regime. We demonstrate this explicitly by evolving the nonlinear scalar field equation in sGB gravity on the Schwarzschild background and observe that indeed a nontrivial scalar field can develop for certain ranges of the parameters if the amplitude of the perturbations is large enough. The obtained in this way scalarized phases are not continuously connected to the Schwarzschild black hole, i.e. they do not bifurcate from it unlike the case of standard scalarization, and they are most probably stable. 

In order to obtain the full spectrum of hairy solutions, including the unstable ones, we solved the fully nonlinear coupled system of reduced field equations. Up to three branches of scalarized phases can exist that have a complicated structure depending on the parameter $\kappa$ in the coupling functions \eqref{eq:coupling_phi4}.  As expected, the stable scalarized phase has the largest entropy among all the branches of hairy black holes. It also has larger entropy that the Schwarzschild phase for most of the parameter range making it thermodynamically preferred. Of course, the entropy only gives an indication about stability and a rigorous study should involve an examination of the linear perturbations that is a study underway.

As a second step we have studied classes of  scalar-Gauss-Bonnet gravity that allow both for linear and nonlinear scalarization. Namely, the coupling function is designed in such a way that the Schwarzschild solution still becomes linearly unstable  below certain mass and thus allow for the appearance of a standard scalarized branch of black holes. As the coupling parameters are varied, though, this scalarized branch is continuously deformed from stable to unstable one. The appearance of such unstable branch is simultaneous with the appears of a new branch of nonlinear scalarized phases and the two are smoothly connected. A range of black holes masses exist where the Schwarzschild solution is linearly stable but still hairy black holes exist where the scalar hair can be excited if a strong nonlinear perturbation is imposed. This completes the picture of fully  nonlinear black hole scalarization in  scalar-Gauss-Bonnet gravity and offers very interesting possibilities for astrophysical manifestations.

The stable scalarized phase is not continuously connected to the Schwarzschild one but instead there is a (sometimes large) ``gap'' between the two. Thus the transition from one phase to the other will happen with a jump contrary to the standard scalarization where the transition from scalarized to non-scalarized state is continuous. This can influence a number of observational phenomena, but let us discuss one particular very interesting example. If we have a nonlinearly scalarized black hole on the border of the maximum mass for the existence of scalarized phases, and this black holes accretes matter, its mass will eventually go beyond the maximum mass for the existence of scalarized phases and it will ``jump'' to the Schwarzschild phase emitting completely its scalar hair. This can happen for example for different X-ray binaries as well as newly former black holes right after a binary neutron star merger or stellar core-collapse. This process does not always require fine tuning of the initial black hole mass because in certain cases, such as the unequal mass neutron star mergers, the mass of the accretion disc can be relatively large \cite{Baiotti:2016qnr,Bernuzzi:2020tgt}. This transition will inevitably lead to strong observational signature, both electromagnetic and gravitational wave, and allow us to perform an independent test of a new sector of sGB gravity.

In the present letter we have examined the time evolution in the decoupling limit approximation, but we are currently performing simulations of stellar-core collapse following \cite{Kuan:2021lol}. These simulations show that the scalar field can be indeed excited nonlinearly in real astrophysical scenarios when the fully nonlinear coupled system of field equations is considered. The specifics of the nonlinear scalarization will be largely dependent also on the matter content of the system and its evolution before the black hole formation. That is why the results will be presented in a follow-up publication.

\textit{Acknowledgements}
DD acknowledges financial support via an Emmy Noether Research Group funded by the German Research Foundation (DFG) under grant
no. DO 1771/1-1.  SY would like to thank the University of Tuebingen for the financial support.  
The partial support by the Bulgarian NSF Grant KP-06-H28/7 and the  Networking support by the COST Actions  CA16104 and CA16214 are also gratefully acknowledged. 

\bibliographystyle{apsrev4-2}
\bibliography{references}

\begin{thebibliography}{38}%
\makeatletter
\providecommand \@ifxundefined [1]{%
 \@ifx{#1\undefined}
}%
\providecommand \@ifnum [1]{%
 \ifnum #1\expandafter \@firstoftwo
 \else \expandafter \@secondoftwo
 \fi
}%
\providecommand \@ifx [1]{%
 \ifx #1\expandafter \@firstoftwo
 \else \expandafter \@secondoftwo
 \fi
}%
\providecommand \natexlab [1]{#1}%
\providecommand \enquote  [1]{``#1''}%
\providecommand \bibnamefont  [1]{#1}%
\providecommand \bibfnamefont [1]{#1}%
\providecommand \citenamefont [1]{#1}%
\providecommand \href@noop [0]{\@secondoftwo}%
\providecommand \href [0]{\begingroup \@sanitize@url \@href}%
\providecommand \@href[1]{\@@startlink{#1}\@@href}%
\providecommand \@@href[1]{\endgroup#1\@@endlink}%
\providecommand \@sanitize@url [0]{\catcode `\\12\catcode `\$12\catcode
  `\&12\catcode `\#12\catcode `\^12\catcode `\_12\catcode `\%12\relax}%
\providecommand \@@startlink[1]{}%
\providecommand \@@endlink[0]{}%
\providecommand \url  [0]{\begingroup\@sanitize@url \@url }%
\providecommand \@url [1]{\endgroup\@href {#1}{\urlprefix }}%
\providecommand \urlprefix  [0]{URL }%
\providecommand \Eprint [0]{\href }%
\providecommand \doibase [0]{https://doi.org/}%
\providecommand \selectlanguage [0]{\@gobble}%
\providecommand \bibinfo  [0]{\@secondoftwo}%
\providecommand \bibfield  [0]{\@secondoftwo}%
\providecommand \translation [1]{[#1]}%
\providecommand \BibitemOpen [0]{}%
\providecommand \bibitemStop [0]{}%
\providecommand \bibitemNoStop [0]{.\EOS\space}%
\providecommand \EOS [0]{\spacefactor3000\relax}%
\providecommand \BibitemShut  [1]{\csname bibitem#1\endcsname}%
\let\auto@bib@innerbib\@empty
\bibitem [{\citenamefont {{Damour}}\ and\ \citenamefont
  {{Esposito-Farese}}(1993)}]{Damour1993}%
  \BibitemOpen
  \bibfield  {author} {\bibinfo {author} {\bibfnamefont {T.}~\bibnamefont
  {{Damour}}}\ and\ \bibinfo {author} {\bibfnamefont {G.}~\bibnamefont
  {{Esposito-Farese}}},\ }\href {https://doi.org/10.1103/PhysRevLett.70.2220}
  {\bibfield  {journal} {\bibinfo  {journal} {Physical Review Letters}\
  }\textbf {\bibinfo {volume} {70}},\ \bibinfo {pages} {2220} (\bibinfo {year}
  {1993})}\BibitemShut {NoStop}%
\bibitem [{\citenamefont {{Damour}}\ and\ \citenamefont
  {{Esposito-Far{\`e}se}}(1996)}]{Damour1996}%
  \BibitemOpen
  \bibfield  {author} {\bibinfo {author} {\bibfnamefont {T.}~\bibnamefont
  {{Damour}}}\ and\ \bibinfo {author} {\bibfnamefont {G.}~\bibnamefont
  {{Esposito-Far{\`e}se}}},\ }\href {https://doi.org/10.1103/PhysRevD.54.1474}
  {\bibfield  {journal} {\bibinfo  {journal} {\prd}\ }\textbf {\bibinfo
  {volume} {54}},\ \bibinfo {pages} {1474} (\bibinfo {year}
  {1996})}\BibitemShut {NoStop}%
\bibitem [{\citenamefont {Doneva}\ and\ \citenamefont
  {Yazadjiev}(2018)}]{Doneva:2017bvd}%
  \BibitemOpen
  \bibfield  {author} {\bibinfo {author} {\bibfnamefont {D.~D.}\ \bibnamefont
  {Doneva}}\ and\ \bibinfo {author} {\bibfnamefont {S.~S.}\ \bibnamefont
  {Yazadjiev}},\ }\href {https://doi.org/10.1103/PhysRevLett.120.131103}
  {\bibfield  {journal} {\bibinfo  {journal} {Phys. Rev. Lett.}\ }\textbf
  {\bibinfo {volume} {120}},\ \bibinfo {pages} {131103} (\bibinfo {year}
  {2018})},\ \Eprint {https://arxiv.org/abs/1711.01187} {arXiv:1711.01187
  [gr-qc]} \BibitemShut {NoStop}%
\bibitem [{\citenamefont {Silva}\ \emph {et~al.}(2018)\citenamefont {Silva},
  \citenamefont {Sakstein}, \citenamefont {Gualtieri}, \citenamefont
  {Sotiriou},\ and\ \citenamefont {Berti}}]{Silva:2017uqg}%
  \BibitemOpen
  \bibfield  {author} {\bibinfo {author} {\bibfnamefont {H.~O.}\ \bibnamefont
  {Silva}}, \bibinfo {author} {\bibfnamefont {J.}~\bibnamefont {Sakstein}},
  \bibinfo {author} {\bibfnamefont {L.}~\bibnamefont {Gualtieri}}, \bibinfo
  {author} {\bibfnamefont {T.~P.}\ \bibnamefont {Sotiriou}},\ and\ \bibinfo
  {author} {\bibfnamefont {E.}~\bibnamefont {Berti}},\ }\href
  {https://doi.org/10.1103/PhysRevLett.120.131104} {\bibfield  {journal}
  {\bibinfo  {journal} {Phys. Rev. Lett.}\ }\textbf {\bibinfo {volume} {120}},\
  \bibinfo {pages} {131104} (\bibinfo {year} {2018})},\ \Eprint
  {https://arxiv.org/abs/1711.02080} {arXiv:1711.02080 [gr-qc]} \BibitemShut
  {NoStop}%
\bibitem [{\citenamefont {Antoniou}\ \emph {et~al.}(2018)\citenamefont
  {Antoniou}, \citenamefont {Bakopoulos},\ and\ \citenamefont
  {Kanti}}]{Antoniou:2017acq}%
  \BibitemOpen
  \bibfield  {author} {\bibinfo {author} {\bibfnamefont {G.}~\bibnamefont
  {Antoniou}}, \bibinfo {author} {\bibfnamefont {A.}~\bibnamefont
  {Bakopoulos}},\ and\ \bibinfo {author} {\bibfnamefont {P.}~\bibnamefont
  {Kanti}},\ }\href {https://doi.org/10.1103/PhysRevLett.120.131102} {\bibfield
   {journal} {\bibinfo  {journal} {Phys. Rev. Lett.}\ }\textbf {\bibinfo
  {volume} {120}},\ \bibinfo {pages} {131102} (\bibinfo {year} {2018})},\
  \Eprint {https://arxiv.org/abs/1711.03390} {arXiv:1711.03390 [hep-th]}
  \BibitemShut {NoStop}%
\bibitem [{\citenamefont {Minamitsuji}\ and\ \citenamefont
  {Ikeda}(2019)}]{Minamitsuji:2018xde}%
  \BibitemOpen
  \bibfield  {author} {\bibinfo {author} {\bibfnamefont {M.}~\bibnamefont
  {Minamitsuji}}\ and\ \bibinfo {author} {\bibfnamefont {T.}~\bibnamefont
  {Ikeda}},\ }\href {https://doi.org/10.1103/PhysRevD.99.044017} {\bibfield
  {journal} {\bibinfo  {journal} {Phys. Rev. D}\ }\textbf {\bibinfo {volume}
  {99}},\ \bibinfo {pages} {044017} (\bibinfo {year} {2019})},\ \Eprint
  {https://arxiv.org/abs/1812.03551} {arXiv:1812.03551 [gr-qc]} \BibitemShut
  {NoStop}%
\bibitem [{\citenamefont {Silva}\ \emph {et~al.}(2019)\citenamefont {Silva},
  \citenamefont {Macedo}, \citenamefont {Sotiriou}, \citenamefont {Gualtieri},
  \citenamefont {Sakstein},\ and\ \citenamefont {Berti}}]{Silva:2018qhn}%
  \BibitemOpen
  \bibfield  {author} {\bibinfo {author} {\bibfnamefont {H.~O.}\ \bibnamefont
  {Silva}}, \bibinfo {author} {\bibfnamefont {C.~F.}\ \bibnamefont {Macedo}},
  \bibinfo {author} {\bibfnamefont {T.~P.}\ \bibnamefont {Sotiriou}}, \bibinfo
  {author} {\bibfnamefont {L.}~\bibnamefont {Gualtieri}}, \bibinfo {author}
  {\bibfnamefont {J.}~\bibnamefont {Sakstein}},\ and\ \bibinfo {author}
  {\bibfnamefont {E.}~\bibnamefont {Berti}},\ }\href
  {https://doi.org/10.1103/PhysRevD.99.064011} {\bibfield  {journal} {\bibinfo
  {journal} {Phys. Rev. D}\ }\textbf {\bibinfo {volume} {99}},\ \bibinfo
  {pages} {064011} (\bibinfo {year} {2019})},\ \Eprint
  {https://arxiv.org/abs/1812.05590} {arXiv:1812.05590 [gr-qc]} \BibitemShut
  {NoStop}%
\bibitem [{\citenamefont {Doneva}\ \emph {et~al.}(2019)\citenamefont {Doneva},
  \citenamefont {Staykov},\ and\ \citenamefont {Yazadjiev}}]{Doneva:2019vuh}%
  \BibitemOpen
  \bibfield  {author} {\bibinfo {author} {\bibfnamefont {D.~D.}\ \bibnamefont
  {Doneva}}, \bibinfo {author} {\bibfnamefont {K.~V.}\ \bibnamefont
  {Staykov}},\ and\ \bibinfo {author} {\bibfnamefont {S.~S.}\ \bibnamefont
  {Yazadjiev}},\ }\href {https://doi.org/10.1103/PhysRevD.99.104045} {\bibfield
   {journal} {\bibinfo  {journal} {Phys. Rev. D}\ }\textbf {\bibinfo {volume}
  {99}},\ \bibinfo {pages} {104045} (\bibinfo {year} {2019})},\ \Eprint
  {https://arxiv.org/abs/1903.08119} {arXiv:1903.08119 [gr-qc]} \BibitemShut
  {NoStop}%
\bibitem [{\citenamefont {Macedo}\ \emph {et~al.}(2019)\citenamefont {Macedo},
  \citenamefont {Sakstein}, \citenamefont {Berti}, \citenamefont {Gualtieri},
  \citenamefont {Silva},\ and\ \citenamefont {Sotiriou}}]{Macedo:2019sem}%
  \BibitemOpen
  \bibfield  {author} {\bibinfo {author} {\bibfnamefont {C.~F.~B.}\
  \bibnamefont {Macedo}}, \bibinfo {author} {\bibfnamefont {J.}~\bibnamefont
  {Sakstein}}, \bibinfo {author} {\bibfnamefont {E.}~\bibnamefont {Berti}},
  \bibinfo {author} {\bibfnamefont {L.}~\bibnamefont {Gualtieri}}, \bibinfo
  {author} {\bibfnamefont {H.~O.}\ \bibnamefont {Silva}},\ and\ \bibinfo
  {author} {\bibfnamefont {T.~P.}\ \bibnamefont {Sotiriou}},\ }\href
  {https://doi.org/10.1103/PhysRevD.99.104041} {\bibfield  {journal} {\bibinfo
  {journal} {Phys. Rev. D}\ }\textbf {\bibinfo {volume} {99}},\ \bibinfo
  {pages} {104041} (\bibinfo {year} {2019})},\ \Eprint
  {https://arxiv.org/abs/1903.06784} {arXiv:1903.06784 [gr-qc]} \BibitemShut
  {NoStop}%
\bibitem [{\citenamefont {Bl\'azquez-Salcedo}\ \emph
  {et~al.}(2021)\citenamefont {Bl\'azquez-Salcedo}, \citenamefont {Kleihaus},\
  and\ \citenamefont {Kunz}}]{Blazquez-Salcedo:2021npn}%
  \BibitemOpen
  \bibfield  {author} {\bibinfo {author} {\bibfnamefont {J.~L.}\ \bibnamefont
  {Bl\'azquez-Salcedo}}, \bibinfo {author} {\bibfnamefont {B.}~\bibnamefont
  {Kleihaus}},\ and\ \bibinfo {author} {\bibfnamefont {J.}~\bibnamefont
  {Kunz}}\ }(\bibinfo {year} {2021})\ \Eprint
  {https://arxiv.org/abs/2106.15574} {arXiv:2106.15574 [gr-qc]} \BibitemShut
  {NoStop}%
\bibitem [{\citenamefont {Antoniou}\ \emph {et~al.}(2021)\citenamefont
  {Antoniou}, \citenamefont {Leh\'ebel}, \citenamefont {Ventagli},\ and\
  \citenamefont {Sotiriou}}]{Antoniou:2021zoy}%
  \BibitemOpen
  \bibfield  {author} {\bibinfo {author} {\bibfnamefont {G.}~\bibnamefont
  {Antoniou}}, \bibinfo {author} {\bibfnamefont {A.}~\bibnamefont {Leh\'ebel}},
  \bibinfo {author} {\bibfnamefont {G.}~\bibnamefont {Ventagli}},\ and\
  \bibinfo {author} {\bibfnamefont {T.~P.}\ \bibnamefont {Sotiriou}},\
  }\href@noop {} {\  (\bibinfo {year} {2021})},\ \Eprint
  {https://arxiv.org/abs/2105.04479} {arXiv:2105.04479 [gr-qc]} \BibitemShut
  {NoStop}%
\bibitem [{\citenamefont {Cunha}\ \emph {et~al.}(2019)\citenamefont {Cunha},
  \citenamefont {Herdeiro},\ and\ \citenamefont {Radu}}]{Cunha:2019dwb}%
  \BibitemOpen
  \bibfield  {author} {\bibinfo {author} {\bibfnamefont {P.~V.}\ \bibnamefont
  {Cunha}}, \bibinfo {author} {\bibfnamefont {C.~A.}\ \bibnamefont
  {Herdeiro}},\ and\ \bibinfo {author} {\bibfnamefont {E.}~\bibnamefont
  {Radu}},\ }\href {https://doi.org/10.1103/PhysRevLett.123.011101} {\bibfield
  {journal} {\bibinfo  {journal} {Phys. Rev. Lett.}\ }\textbf {\bibinfo
  {volume} {123}},\ \bibinfo {pages} {011101} (\bibinfo {year} {2019})},\
  \Eprint {https://arxiv.org/abs/1904.09997} {arXiv:1904.09997 [gr-qc]}
  \BibitemShut {NoStop}%
\bibitem [{\citenamefont {Collodel}\ \emph {et~al.}(2020)\citenamefont
  {Collodel}, \citenamefont {Kleihaus}, \citenamefont {Kunz},\ and\
  \citenamefont {Berti}}]{Collodel:2019kkx}%
  \BibitemOpen
  \bibfield  {author} {\bibinfo {author} {\bibfnamefont {L.~G.}\ \bibnamefont
  {Collodel}}, \bibinfo {author} {\bibfnamefont {B.}~\bibnamefont {Kleihaus}},
  \bibinfo {author} {\bibfnamefont {J.}~\bibnamefont {Kunz}},\ and\ \bibinfo
  {author} {\bibfnamefont {E.}~\bibnamefont {Berti}},\ }\href
  {https://doi.org/10.1088/1361-6382/ab74f9} {\bibfield  {journal} {\bibinfo
  {journal} {Class. Quant. Grav.}\ }\textbf {\bibinfo {volume} {37}},\ \bibinfo
  {pages} {075018} (\bibinfo {year} {2020})},\ \Eprint
  {https://arxiv.org/abs/1912.05382} {arXiv:1912.05382 [gr-qc]} \BibitemShut
  {NoStop}%
\bibitem [{\citenamefont {Dima}\ \emph {et~al.}(2020)\citenamefont {Dima},
  \citenamefont {Barausse}, \citenamefont {Franchini},\ and\ \citenamefont
  {Sotiriou}}]{Dima:2020yac}%
  \BibitemOpen
  \bibfield  {author} {\bibinfo {author} {\bibfnamefont {A.}~\bibnamefont
  {Dima}}, \bibinfo {author} {\bibfnamefont {E.}~\bibnamefont {Barausse}},
  \bibinfo {author} {\bibfnamefont {N.}~\bibnamefont {Franchini}},\ and\
  \bibinfo {author} {\bibfnamefont {T.~P.}\ \bibnamefont {Sotiriou}},\ }\href
  {https://doi.org/10.1103/PhysRevLett.125.231101} {\bibfield  {journal}
  {\bibinfo  {journal} {Phys. Rev. Lett.}\ }\textbf {\bibinfo {volume} {125}},\
  \bibinfo {pages} {231101} (\bibinfo {year} {2020})},\ \Eprint
  {https://arxiv.org/abs/2006.03095} {arXiv:2006.03095 [gr-qc]} \BibitemShut
  {NoStop}%
\bibitem [{\citenamefont {Doneva}\ \emph
  {et~al.}(2020{\natexlab{a}})\citenamefont {Doneva}, \citenamefont {Collodel},
  \citenamefont {Kr\"uger},\ and\ \citenamefont {Yazadjiev}}]{Doneva:2020nbb}%
  \BibitemOpen
  \bibfield  {author} {\bibinfo {author} {\bibfnamefont {D.~D.}\ \bibnamefont
  {Doneva}}, \bibinfo {author} {\bibfnamefont {L.~G.}\ \bibnamefont
  {Collodel}}, \bibinfo {author} {\bibfnamefont {C.~J.}\ \bibnamefont
  {Kr\"uger}},\ and\ \bibinfo {author} {\bibfnamefont {S.~S.}\ \bibnamefont
  {Yazadjiev}},\ }\href {https://doi.org/10.1103/PhysRevD.102.104027}
  {\bibfield  {journal} {\bibinfo  {journal} {Phys. Rev. D}\ }\textbf {\bibinfo
  {volume} {102}},\ \bibinfo {pages} {104027} (\bibinfo {year}
  {2020}{\natexlab{a}})},\ \Eprint {https://arxiv.org/abs/2008.07391}
  {arXiv:2008.07391 [gr-qc]} \BibitemShut {NoStop}%
\bibitem [{\citenamefont {Doneva}\ \emph
  {et~al.}(2020{\natexlab{b}})\citenamefont {Doneva}, \citenamefont {Collodel},
  \citenamefont {Kr\"uger},\ and\ \citenamefont {Yazadjiev}}]{Doneva:2020kfv}%
  \BibitemOpen
  \bibfield  {author} {\bibinfo {author} {\bibfnamefont {D.~D.}\ \bibnamefont
  {Doneva}}, \bibinfo {author} {\bibfnamefont {L.~G.}\ \bibnamefont
  {Collodel}}, \bibinfo {author} {\bibfnamefont {C.~J.}\ \bibnamefont
  {Kr\"uger}},\ and\ \bibinfo {author} {\bibfnamefont {S.~S.}\ \bibnamefont
  {Yazadjiev}},\ }\href@noop {} {\  (\bibinfo {year} {2020}{\natexlab{b}})},\
  \Eprint {https://arxiv.org/abs/2009.03774} {arXiv:2009.03774 [gr-qc]}
  \BibitemShut {NoStop}%
\bibitem [{\citenamefont {Herdeiro}\ \emph {et~al.}(2020)\citenamefont
  {Herdeiro}, \citenamefont {Radu}, \citenamefont {Silva}, \citenamefont
  {Sotiriou},\ and\ \citenamefont {Yunes}}]{Herdeiro:2020wei}%
  \BibitemOpen
  \bibfield  {author} {\bibinfo {author} {\bibfnamefont {C.~A.}\ \bibnamefont
  {Herdeiro}}, \bibinfo {author} {\bibfnamefont {E.}~\bibnamefont {Radu}},
  \bibinfo {author} {\bibfnamefont {H.~O.}\ \bibnamefont {Silva}}, \bibinfo
  {author} {\bibfnamefont {T.~P.}\ \bibnamefont {Sotiriou}},\ and\ \bibinfo
  {author} {\bibfnamefont {N.}~\bibnamefont {Yunes}},\ }\href@noop {} {\
  (\bibinfo {year} {2020})},\ \Eprint {https://arxiv.org/abs/2009.03904}
  {arXiv:2009.03904 [gr-qc]} \BibitemShut {NoStop}%
\bibitem [{\citenamefont {Berti}\ \emph {et~al.}(2020)\citenamefont {Berti},
  \citenamefont {Collodel}, \citenamefont {Kleihaus},\ and\ \citenamefont
  {Kunz}}]{Berti:2020kgk}%
  \BibitemOpen
  \bibfield  {author} {\bibinfo {author} {\bibfnamefont {E.}~\bibnamefont
  {Berti}}, \bibinfo {author} {\bibfnamefont {L.~G.}\ \bibnamefont {Collodel}},
  \bibinfo {author} {\bibfnamefont {B.}~\bibnamefont {Kleihaus}},\ and\
  \bibinfo {author} {\bibfnamefont {J.}~\bibnamefont {Kunz}},\ }\href@noop {}
  {\  (\bibinfo {year} {2020})},\ \Eprint {https://arxiv.org/abs/2009.03905}
  {arXiv:2009.03905 [gr-qc]} \BibitemShut {NoStop}%
\bibitem [{\citenamefont {Bl\'azquez-Salcedo}\ \emph
  {et~al.}(2018)\citenamefont {Bl\'azquez-Salcedo}, \citenamefont {Doneva},
  \citenamefont {Kunz},\ and\ \citenamefont
  {Yazadjiev}}]{Blazquez-Salcedo:2018jnn}%
  \BibitemOpen
  \bibfield  {author} {\bibinfo {author} {\bibfnamefont {J.~L.}\ \bibnamefont
  {Bl\'azquez-Salcedo}}, \bibinfo {author} {\bibfnamefont {D.~D.}\ \bibnamefont
  {Doneva}}, \bibinfo {author} {\bibfnamefont {J.}~\bibnamefont {Kunz}},\ and\
  \bibinfo {author} {\bibfnamefont {S.~S.}\ \bibnamefont {Yazadjiev}},\ }\href
  {https://doi.org/10.1103/PhysRevD.98.084011} {\bibfield  {journal} {\bibinfo
  {journal} {Phys. Rev. D}\ }\textbf {\bibinfo {volume} {98}},\ \bibinfo
  {pages} {084011} (\bibinfo {year} {2018})},\ \Eprint
  {https://arxiv.org/abs/1805.05755} {arXiv:1805.05755 [gr-qc]} \BibitemShut
  {NoStop}%
\bibitem [{\citenamefont {Bl\'azquez-Salcedo}\ \emph
  {et~al.}(2020{\natexlab{a}})\citenamefont {Bl\'azquez-Salcedo}, \citenamefont
  {Doneva}, \citenamefont {Kahlen}, \citenamefont {Kunz}, \citenamefont
  {Nedkova},\ and\ \citenamefont {Yazadjiev}}]{Blazquez-Salcedo:2020rhf}%
  \BibitemOpen
  \bibfield  {author} {\bibinfo {author} {\bibfnamefont {J.~L.}\ \bibnamefont
  {Bl\'azquez-Salcedo}}, \bibinfo {author} {\bibfnamefont {D.~D.}\ \bibnamefont
  {Doneva}}, \bibinfo {author} {\bibfnamefont {S.}~\bibnamefont {Kahlen}},
  \bibinfo {author} {\bibfnamefont {J.}~\bibnamefont {Kunz}}, \bibinfo {author}
  {\bibfnamefont {P.}~\bibnamefont {Nedkova}},\ and\ \bibinfo {author}
  {\bibfnamefont {S.~S.}\ \bibnamefont {Yazadjiev}},\ }\href
  {https://doi.org/10.1103/PhysRevD.101.104006} {\bibfield  {journal} {\bibinfo
   {journal} {Phys. Rev. D}\ }\textbf {\bibinfo {volume} {101}},\ \bibinfo
  {pages} {104006} (\bibinfo {year} {2020}{\natexlab{a}})},\ \Eprint
  {https://arxiv.org/abs/2003.02862} {arXiv:2003.02862 [gr-qc]} \BibitemShut
  {NoStop}%
\bibitem [{\citenamefont {Bl\'azquez-Salcedo}\ \emph
  {et~al.}(2020{\natexlab{b}})\citenamefont {Bl\'azquez-Salcedo}, \citenamefont
  {Doneva}, \citenamefont {Kahlen}, \citenamefont {Kunz}, \citenamefont
  {Nedkova},\ and\ \citenamefont {Yazadjiev}}]{Blazquez-Salcedo:2020caw}%
  \BibitemOpen
  \bibfield  {author} {\bibinfo {author} {\bibfnamefont {J.~L.}\ \bibnamefont
  {Bl\'azquez-Salcedo}}, \bibinfo {author} {\bibfnamefont {D.~D.}\ \bibnamefont
  {Doneva}}, \bibinfo {author} {\bibfnamefont {S.}~\bibnamefont {Kahlen}},
  \bibinfo {author} {\bibfnamefont {J.}~\bibnamefont {Kunz}}, \bibinfo {author}
  {\bibfnamefont {P.}~\bibnamefont {Nedkova}},\ and\ \bibinfo {author}
  {\bibfnamefont {S.~S.}\ \bibnamefont {Yazadjiev}},\ }\href
  {https://doi.org/10.1103/PhysRevD.102.024086} {\bibfield  {journal} {\bibinfo
   {journal} {Phys. Rev. D}\ }\textbf {\bibinfo {volume} {102}},\ \bibinfo
  {pages} {024086} (\bibinfo {year} {2020}{\natexlab{b}})},\ \Eprint
  {https://arxiv.org/abs/2006.06006} {arXiv:2006.06006 [gr-qc]} \BibitemShut
  {NoStop}%
\bibitem [{\citenamefont {Ripley}\ and\ \citenamefont
  {Pretorius}(2020)}]{Ripley:2020vpk}%
  \BibitemOpen
  \bibfield  {author} {\bibinfo {author} {\bibfnamefont {J.~L.}\ \bibnamefont
  {Ripley}}\ and\ \bibinfo {author} {\bibfnamefont {F.}~\bibnamefont
  {Pretorius}},\ }\href {https://doi.org/10.1088/1361-6382/ab9bbb} {\bibfield
  {journal} {\bibinfo  {journal} {Class. Quant. Grav.}\ }\textbf {\bibinfo
  {volume} {37}},\ \bibinfo {pages} {155003} (\bibinfo {year} {2020})},\
  \Eprint {https://arxiv.org/abs/2005.05417} {arXiv:2005.05417 [gr-qc]}
  \BibitemShut {NoStop}%
\bibitem [{\citenamefont {Silva}\ \emph {et~al.}(2020)\citenamefont {Silva},
  \citenamefont {Witek}, \citenamefont {Elley},\ and\ \citenamefont
  {Yunes}}]{Silva:2020omi}%
  \BibitemOpen
  \bibfield  {author} {\bibinfo {author} {\bibfnamefont {H.~O.}\ \bibnamefont
  {Silva}}, \bibinfo {author} {\bibfnamefont {H.}~\bibnamefont {Witek}},
  \bibinfo {author} {\bibfnamefont {M.}~\bibnamefont {Elley}},\ and\ \bibinfo
  {author} {\bibfnamefont {N.}~\bibnamefont {Yunes}},\ }\href@noop {} {\
  (\bibinfo {year} {2020})},\ \Eprint {https://arxiv.org/abs/2012.10436}
  {arXiv:2012.10436 [gr-qc]} \BibitemShut {NoStop}%
\bibitem [{\citenamefont {Doneva}\ and\ \citenamefont
  {Yazadjiev}(2021{\natexlab{a}})}]{Doneva:2021dqn}%
  \BibitemOpen
  \bibfield  {author} {\bibinfo {author} {\bibfnamefont {D.~D.}\ \bibnamefont
  {Doneva}}\ and\ \bibinfo {author} {\bibfnamefont {S.~S.}\ \bibnamefont
  {Yazadjiev}},\ }\href {https://doi.org/10.1103/PhysRevD.103.064024}
  {\bibfield  {journal} {\bibinfo  {journal} {Phys. Rev. D}\ }\textbf {\bibinfo
  {volume} {103}},\ \bibinfo {pages} {064024} (\bibinfo {year}
  {2021}{\natexlab{a}})},\ \Eprint {https://arxiv.org/abs/2101.03514}
  {arXiv:2101.03514 [gr-qc]} \BibitemShut {NoStop}%
\bibitem [{\citenamefont {Kuan}\ \emph {et~al.}(2021)\citenamefont {Kuan},
  \citenamefont {Doneva},\ and\ \citenamefont {Yazadjiev}}]{Kuan:2021lol}%
  \BibitemOpen
  \bibfield  {author} {\bibinfo {author} {\bibfnamefont {H.-J.}\ \bibnamefont
  {Kuan}}, \bibinfo {author} {\bibfnamefont {D.~D.}\ \bibnamefont {Doneva}},\
  and\ \bibinfo {author} {\bibfnamefont {S.~S.}\ \bibnamefont {Yazadjiev}},\
  }\href {https://doi.org/10.1103/PhysRevLett.127.161103} {\bibfield  {journal}
  {\bibinfo  {journal} {Phys. Rev. Lett.}\ }\textbf {\bibinfo {volume} {127}},\
  \bibinfo {pages} {161103} (\bibinfo {year} {2021})},\ \Eprint
  {https://arxiv.org/abs/2103.11999} {arXiv:2103.11999 [gr-qc]} \BibitemShut
  {NoStop}%
\bibitem [{\citenamefont {East}\ and\ \citenamefont
  {Ripley}(2021)}]{East:2021bqk}%
  \BibitemOpen
  \bibfield  {author} {\bibinfo {author} {\bibfnamefont {W.~E.}\ \bibnamefont
  {East}}\ and\ \bibinfo {author} {\bibfnamefont {J.~L.}\ \bibnamefont
  {Ripley}},\ }\href@noop {} {\  (\bibinfo {year} {2021})},\ \Eprint
  {https://arxiv.org/abs/2105.08571} {arXiv:2105.08571 [gr-qc]} \BibitemShut
  {NoStop}%
\bibitem [{\citenamefont {Herdeiro}\ \emph {et~al.}(2018)\citenamefont
  {Herdeiro}, \citenamefont {Radu}, \citenamefont {Sanchis-Gual},\ and\
  \citenamefont {Font}}]{Herdeiro:2018wub}%
  \BibitemOpen
  \bibfield  {author} {\bibinfo {author} {\bibfnamefont {C.~A.}\ \bibnamefont
  {Herdeiro}}, \bibinfo {author} {\bibfnamefont {E.}~\bibnamefont {Radu}},
  \bibinfo {author} {\bibfnamefont {N.}~\bibnamefont {Sanchis-Gual}},\ and\
  \bibinfo {author} {\bibfnamefont {J.~A.}\ \bibnamefont {Font}},\ }\href
  {https://doi.org/10.1103/PhysRevLett.121.101102} {\bibfield  {journal}
  {\bibinfo  {journal} {Phys. Rev. Lett.}\ }\textbf {\bibinfo {volume} {121}},\
  \bibinfo {pages} {101102} (\bibinfo {year} {2018})},\ \Eprint
  {https://arxiv.org/abs/1806.05190} {arXiv:1806.05190 [gr-qc]} \BibitemShut
  {NoStop}%
\bibitem [{\citenamefont {Andreou}\ \emph {et~al.}(2019)\citenamefont
  {Andreou}, \citenamefont {Franchini}, \citenamefont {Ventagli},\ and\
  \citenamefont {Sotiriou}}]{Andreou:2019ikc}%
  \BibitemOpen
  \bibfield  {author} {\bibinfo {author} {\bibfnamefont {N.}~\bibnamefont
  {Andreou}}, \bibinfo {author} {\bibfnamefont {N.}~\bibnamefont {Franchini}},
  \bibinfo {author} {\bibfnamefont {G.}~\bibnamefont {Ventagli}},\ and\
  \bibinfo {author} {\bibfnamefont {T.~P.}\ \bibnamefont {Sotiriou}},\ }\href
  {https://doi.org/10.1103/PhysRevD.99.124022} {\bibfield  {journal} {\bibinfo
  {journal} {Phys. Rev. D}\ }\textbf {\bibinfo {volume} {99}},\ \bibinfo
  {pages} {124022} (\bibinfo {year} {2019})},\ \bibinfo {note} {[Erratum:
  Phys.Rev.D 101, 109903 (2020)]},\ \Eprint {https://arxiv.org/abs/1904.06365}
  {arXiv:1904.06365 [gr-qc]} \BibitemShut {NoStop}%
\bibitem [{\citenamefont {Gao}\ \emph {et~al.}(2019)\citenamefont {Gao},
  \citenamefont {Huang},\ and\ \citenamefont {Liu}}]{Gao:2018acg}%
  \BibitemOpen
  \bibfield  {author} {\bibinfo {author} {\bibfnamefont {Y.-X.}\ \bibnamefont
  {Gao}}, \bibinfo {author} {\bibfnamefont {Y.}~\bibnamefont {Huang}},\ and\
  \bibinfo {author} {\bibfnamefont {D.-J.}\ \bibnamefont {Liu}},\ }\href
  {https://doi.org/10.1103/PhysRevD.99.044020} {\bibfield  {journal} {\bibinfo
  {journal} {Phys. Rev. D}\ }\textbf {\bibinfo {volume} {99}},\ \bibinfo
  {pages} {044020} (\bibinfo {year} {2019})},\ \Eprint
  {https://arxiv.org/abs/1808.01433} {arXiv:1808.01433 [gr-qc]} \BibitemShut
  {NoStop}%
\bibitem [{\citenamefont {Doneva}\ and\ \citenamefont
  {Yazadjiev}(2021{\natexlab{b}})}]{Doneva:2021dcc}%
  \BibitemOpen
  \bibfield  {author} {\bibinfo {author} {\bibfnamefont {D.~D.}\ \bibnamefont
  {Doneva}}\ and\ \bibinfo {author} {\bibfnamefont {S.~S.}\ \bibnamefont
  {Yazadjiev}},\ }\href {https://doi.org/10.1103/PhysRevD.103.083007}
  {\bibfield  {journal} {\bibinfo  {journal} {Phys. Rev. D}\ }\textbf {\bibinfo
  {volume} {103}},\ \bibinfo {pages} {083007} (\bibinfo {year}
  {2021}{\natexlab{b}})},\ \Eprint {https://arxiv.org/abs/2102.03940}
  {arXiv:2102.03940 [gr-qc]} \BibitemShut {NoStop}%
\bibitem [{\citenamefont {Zhang}(2021)}]{Zhang:2021btn}%
  \BibitemOpen
  \bibfield  {author} {\bibinfo {author} {\bibfnamefont {S.-J.}\ \bibnamefont
  {Zhang}},\ }\href {https://doi.org/10.1140/epjc/s10052-021-09249-8}
  {\bibfield  {journal} {\bibinfo  {journal} {Eur. Phys. J. C}\ }\textbf
  {\bibinfo {volume} {81}},\ \bibinfo {pages} {441} (\bibinfo {year} {2021})},\
  \Eprint {https://arxiv.org/abs/2102.10479} {arXiv:2102.10479 [gr-qc]}
  \BibitemShut {NoStop}%
\bibitem [{\citenamefont {Myung}\ and\ \citenamefont
  {Zou}(2021)}]{Myung:2021ztl}%
  \BibitemOpen
  \bibfield  {author} {\bibinfo {author} {\bibfnamefont {Y.~S.}\ \bibnamefont
  {Myung}}\ and\ \bibinfo {author} {\bibfnamefont {D.-C.}\ \bibnamefont
  {Zou}},\ }\href@noop {} {\  (\bibinfo {year} {2021})},\ \Eprint
  {https://arxiv.org/abs/2103.01389} {arXiv:2103.01389 [gr-qc]} \BibitemShut
  {NoStop}%
\bibitem [{\citenamefont {Bl\'azquez-Salcedo}\ \emph
  {et~al.}(2020{\natexlab{c}})\citenamefont {Bl\'azquez-Salcedo}, \citenamefont
  {Herdeiro}, \citenamefont {Kunz}, \citenamefont {Pombo},\ and\ \citenamefont
  {Radu}}]{Blazquez-Salcedo:2020nhs}%
  \BibitemOpen
  \bibfield  {author} {\bibinfo {author} {\bibfnamefont {J.~L.}\ \bibnamefont
  {Bl\'azquez-Salcedo}}, \bibinfo {author} {\bibfnamefont {C.~A.~R.}\
  \bibnamefont {Herdeiro}}, \bibinfo {author} {\bibfnamefont {J.}~\bibnamefont
  {Kunz}}, \bibinfo {author} {\bibfnamefont {A.~M.}\ \bibnamefont {Pombo}},\
  and\ \bibinfo {author} {\bibfnamefont {E.}~\bibnamefont {Radu}},\ }\href
  {https://doi.org/10.1016/j.physletb.2020.135493} {\bibfield  {journal}
  {\bibinfo  {journal} {Phys. Lett. B}\ }\textbf {\bibinfo {volume} {806}},\
  \bibinfo {pages} {135493} (\bibinfo {year} {2020}{\natexlab{c}})},\ \Eprint
  {https://arxiv.org/abs/2002.00963} {arXiv:2002.00963 [gr-qc]} \BibitemShut
  {NoStop}%
\bibitem [{\citenamefont {Kleihaus}\ \emph
  {et~al.}(2020{\natexlab{a}})\citenamefont {Kleihaus}, \citenamefont {Kunz},\
  and\ \citenamefont {Kanti}}]{Kleihaus:2019rbg}%
  \BibitemOpen
  \bibfield  {author} {\bibinfo {author} {\bibfnamefont {B.}~\bibnamefont
  {Kleihaus}}, \bibinfo {author} {\bibfnamefont {J.}~\bibnamefont {Kunz}},\
  and\ \bibinfo {author} {\bibfnamefont {P.}~\bibnamefont {Kanti}},\ }\href
  {https://doi.org/10.1016/j.physletb.2020.135401} {\bibfield  {journal}
  {\bibinfo  {journal} {Phys. Lett. B}\ }\textbf {\bibinfo {volume} {804}},\
  \bibinfo {pages} {135401} (\bibinfo {year} {2020}{\natexlab{a}})},\ \Eprint
  {https://arxiv.org/abs/1910.02121} {arXiv:1910.02121 [gr-qc]} \BibitemShut
  {NoStop}%
\bibitem [{\citenamefont {Kleihaus}\ \emph
  {et~al.}(2020{\natexlab{b}})\citenamefont {Kleihaus}, \citenamefont {Kunz},\
  and\ \citenamefont {Kanti}}]{Kleihaus:2020qwo}%
  \BibitemOpen
  \bibfield  {author} {\bibinfo {author} {\bibfnamefont {B.}~\bibnamefont
  {Kleihaus}}, \bibinfo {author} {\bibfnamefont {J.}~\bibnamefont {Kunz}},\
  and\ \bibinfo {author} {\bibfnamefont {P.}~\bibnamefont {Kanti}},\ }\href
  {https://doi.org/10.1103/PhysRevD.102.024070} {\bibfield  {journal} {\bibinfo
   {journal} {Phys. Rev. D}\ }\textbf {\bibinfo {volume} {102}},\ \bibinfo
  {pages} {024070} (\bibinfo {year} {2020}{\natexlab{b}})},\ \Eprint
  {https://arxiv.org/abs/2005.07650} {arXiv:2005.07650 [gr-qc]} \BibitemShut
  {NoStop}%
\bibitem [{\citenamefont {Doneva}\ \emph {et~al.}(2018)\citenamefont {Doneva},
  \citenamefont {Kiorpelidi}, \citenamefont {Nedkova}, \citenamefont
  {Papantonopoulos},\ and\ \citenamefont {Yazadjiev}}]{Doneva:2018rou}%
  \BibitemOpen
  \bibfield  {author} {\bibinfo {author} {\bibfnamefont {D.~D.}\ \bibnamefont
  {Doneva}}, \bibinfo {author} {\bibfnamefont {S.}~\bibnamefont {Kiorpelidi}},
  \bibinfo {author} {\bibfnamefont {P.~G.}\ \bibnamefont {Nedkova}}, \bibinfo
  {author} {\bibfnamefont {E.}~\bibnamefont {Papantonopoulos}},\ and\ \bibinfo
  {author} {\bibfnamefont {S.~S.}\ \bibnamefont {Yazadjiev}},\ }\href
  {https://doi.org/10.1103/PhysRevD.98.104056} {\bibfield  {journal} {\bibinfo
  {journal} {Phys. Rev. D}\ }\textbf {\bibinfo {volume} {98}},\ \bibinfo
  {pages} {104056} (\bibinfo {year} {2018})},\ \Eprint
  {https://arxiv.org/abs/1809.00844} {arXiv:1809.00844 [gr-qc]} \BibitemShut
  {NoStop}%
\bibitem [{\citenamefont {Baiotti}\ and\ \citenamefont
  {Rezzolla}(2017)}]{Baiotti:2016qnr}%
  \BibitemOpen
  \bibfield  {author} {\bibinfo {author} {\bibfnamefont {L.}~\bibnamefont
  {Baiotti}}\ and\ \bibinfo {author} {\bibfnamefont {L.}~\bibnamefont
  {Rezzolla}},\ }\href {https://doi.org/10.1088/1361-6633/aa67bb} {\bibfield
  {journal} {\bibinfo  {journal} {Rept. Prog. Phys.}\ }\textbf {\bibinfo
  {volume} {80}},\ \bibinfo {pages} {096901} (\bibinfo {year} {2017})},\
  \Eprint {https://arxiv.org/abs/1607.03540} {arXiv:1607.03540 [gr-qc]}
  \BibitemShut {NoStop}%
\bibitem [{\citenamefont {Bernuzzi}(2020)}]{Bernuzzi:2020tgt}%
  \BibitemOpen
  \bibfield  {author} {\bibinfo {author} {\bibfnamefont {S.}~\bibnamefont
  {Bernuzzi}},\ }\href {https://doi.org/10.1007/s10714-020-02752-5} {\bibfield
  {journal} {\bibinfo  {journal} {Gen. Rel. Grav.}\ }\textbf {\bibinfo {volume}
  {52}},\ \bibinfo {pages} {108} (\bibinfo {year} {2020})},\ \Eprint
  {https://arxiv.org/abs/2004.06419} {arXiv:2004.06419 [astro-ph.HE]}
  \BibitemShut {NoStop}%
\end{thebibliography}%

\section{Supplemental material}

\subsection{Basic mathematical equations} \label{app1}

The variation of the action in scalar-Gauss-Bonnet (sGB) gravity  with respect to the metric $g_{\mu\nu}$ and scalar field $\varphi$ gives the following field equations 

\begin{eqnarray}\label{FE}
	&&R_{\mu\nu}- \frac{1}{2}R g_{\mu\nu} + \Gamma_{\mu\nu}= 2\nabla_\mu\varphi\nabla_\nu\varphi -  g_{\mu\nu} \nabla_\alpha\varphi \nabla^\alpha\varphi ,\\
	&&\nabla_\alpha\nabla^\alpha\varphi=  -  \frac{\lambda^2}{4} \frac{df(\varphi)}{d\varphi} {\cal R}^2_{GB},\label{SFE}
\end{eqnarray}
where  $\nabla_{\mu}$ is the covariant derivative with respect to  $g_{\mu\nu}$ and  $\Gamma_{\mu\nu}$ is defined by 
\begin{eqnarray}
	\Gamma_{\mu\nu} &=& - R(\nabla_\mu\Psi_{\nu} + \nabla_\nu\Psi_{\mu} ) - 4\nabla^\alpha\Psi_{\alpha}\left(R_{\mu\nu} - \frac{1}{2}R g_{\mu\nu}\right) \nonumber \\
	& +& 4R_{\mu\alpha}\nabla^\alpha\Psi_{\nu} + 4R_{\nu\alpha}\nabla^\alpha\Psi_{\mu} \nonumber \\ 
	& - & 4 g_{\mu\nu} R^{\alpha\beta}\nabla_\alpha\Psi_{\beta} 
	+ \,  4 R^{\beta}_{\;\mu\alpha\nu}\nabla^\alpha\Psi_{\beta} 
\end{eqnarray}  
with 
\begin{eqnarray}
	\Psi_{\mu}= \lambda^2 \frac{df(\varphi)}{d\varphi}\nabla_\mu\varphi .
\end{eqnarray}

\subsubsection{Equations governing the dynamics}	

In the decoupling limit we solve eq. \eqref{SFE} on the Schwarzschild background which in the standard coordinate reads 
\begin{eqnarray}
	ds^2= - \frac{\Delta}{r^2} dt^2 + \frac{r^2}{\Delta} dr^2 + r^2 (d\theta^2 + \sin^2\theta d\phi^2),
\end{eqnarray} 
where $\Delta=r^2-2Mr$. It is convenient to introduce the coordinate $x$ defined by $dx=\frac{r^2}{\Delta} dr$. In the coordinates 
$(t, x, \theta,\phi)$ eq. \eqref{SFE} takes the following explicit form
\begin{eqnarray}\label{eq:TimeEvolEq}
	&&- \partial^2_t \varphi + \partial^2_x \varphi  + \frac{2 \Delta}{r^3} \partial_x\delta\varphi  \\ 
	&&+ \frac{\Delta}{r^4}\left[\frac{1}{\sin\theta} \partial_\theta(\sin\theta\partial_\theta\varphi) +   \frac{1}{\sin^2\theta}\partial^2_{\phi}\varphi \right] 
	= -\lambda^2 \frac{12 M^2\Delta}{r^8}\frac{df(\varphi)}{d\varphi}. \nonumber
\end{eqnarray}

The boundary conditions we have to impose when evolving in time eq. \eqref{eq:TimeEvolEq} is that the scalar
field  has the form of an outgoing wave at infinity and an ingoing wave at the black
hole horizon

\subsubsection{Static and spherically symmetric equations} 

We consider further static and spherically symmetric spacetimes as well as static and spherically symmetric scalar field configurations. The spacetime metric can be written then as
\begin{eqnarray}
	ds^2= - e^{2\Phi(r)}dt^2 + e^{2\Lambda(r)} dr^2 + r^2 (d\theta^2 + \sin^2\theta d\phi^2 ). 
\end{eqnarray}   
After using this form  of the metric a system of reduced field equations can be derived that describes the static black hole solutions in sGB gravity
\begin{eqnarray}
	&&\frac{2}{r}\left[1 +  \frac{2}{r} (1-3e^{-2\Lambda})  \Psi_{r}  \right]  \frac{d\Lambda}{dr} + \frac{(e^{2\Lambda}-1)}{r^2} \nonumber \\
	&&	\hspace{0.5cm} - \frac{4}{r^2}(1-e^{-2\Lambda}) \frac{d\Psi_{r}}{dr} - \left( \frac{d\varphi}{dr}\right)^2=0, \label{DRFE1}\\ && \nonumber \\
	&&\frac{2}{r}\left[1 +  \frac{2}{r} (1-3e^{-2\Lambda})  \Psi_{r}  \right]  \frac{d\Phi}{dr} - \frac{(e^{2\Lambda}-1)}{r^2} - \left( \frac{d\varphi}{dr}\right)^2=0,\label{DRFE2}\\ && \nonumber \\
	&& \frac{d^2\Phi}{dr^2} + \left(\frac{d\Phi}{dr} + \frac{1}{r}\right)\left(\frac{d\Phi}{dr} - \frac{d\Lambda}{dr}\right)  \nonumber \\
	&& \hspace{0.5cm} + \frac{4e^{-2\Lambda}}{r}\left[3\frac{d\Phi}{dr}\frac{d\Lambda}{dr} - \frac{d^2\Phi}{dr^2} - \left(\frac{d\Phi}{dr}\right)^2 \right]\Psi_{r} 
	\nonumber \\ 
	&& \hspace{0.5cm} - \frac{4e^{-2\Lambda}}{r}\frac{d\Phi}{dr} \frac{d\Psi_r}{dr} + \left(\frac{d\varphi}{dr}\right)^2=0, \label{DRFE3}\\ && \nonumber \\
	&& \frac{d^2\varphi}{dr^2}  + \left(\frac{d\Phi}{dr} \nonumber - \frac{d\Lambda}{dr} + \frac{2}{r}\right)\frac{d\varphi}{dr} \nonumber \\ 
	&& \hspace{0.5cm} - \frac{2\lambda^2}{r^2} \frac{df(\varphi)}{d\phi}\Big\{(1-e^{-2\Lambda})\left[\frac{d^2\Phi}{dr^2} + \frac{d\Phi}{dr} \left(\frac{d\Phi}{dr} - \frac{d\Lambda}{dr}\right)\right]   \nonumber \\
	&&  \hspace{0.5cm} + 2e^{-2\Lambda}\frac{d\Phi}{dr} \frac{d\Lambda}{dr}\Big\} =0, \label{DRFE4}
\end{eqnarray}
with 
\begin{eqnarray}
	\Psi_{r}=\lambda^2 \frac{df(\varphi)}{d\varphi} \frac{d\varphi}{dr}.
\end{eqnarray}
In order to obtain black hole solutions the following conditions should be imposed coming from the requirements for asymptotic flatness at infinity and the regularity at the black hole horizon $r=r_H$:
\begin{eqnarray}
	&&\Phi|_{r\rightarrow\infty} \rightarrow 0, \;\;  \Lambda|_{r\rightarrow\infty} \rightarrow 0,\;\; \varphi|_{r\rightarrow\infty} \rightarrow 0\;\;,   \label{eq:BH_inf} \\
	&&e^{2\Phi}|_{r\rightarrow r_H} \rightarrow 0, \;\; e^{-2\Lambda}|_{r\rightarrow r_H} \rightarrow 0. \label{eq:BC_rh}
\end{eqnarray} 
The regularity of the scalar field and its first and second derivatives on the black hole horizon leads to an additional condition for the existence of black hole solutions
\begin{eqnarray}
	\left(\frac{d\varphi}{dr}\right)_{H}&=& \frac{r_{H}}{4 \lambda^2 \frac{df}{d\varphi}(\varphi_{H})}  \times \nonumber \\
	&&\times \left[-1 + \sqrt{1 - \frac{24\lambda^4}{r^4_{H}} \left(\frac{df}{d\varphi}(\varphi_{H})\right)^2}\right]. \label{eq:RegularityCond}
\end{eqnarray} 
Clearly, solutions can exist only in case the expression inside the square root is positive. For black holes with nontrivial scalar field this condition can be easily violated, though, and it limits (sometimes severely) the domain of existence of scalarized solutions \cite{Doneva:2017bvd}.

\subsection{Numerical results for the second coupling function $f_2(\varphi)$} \label{app2}

\subsubsection{Dynamics with $f_2(\varphi)$}
\begin{figure}
	\includegraphics[width=0.45\textwidth]{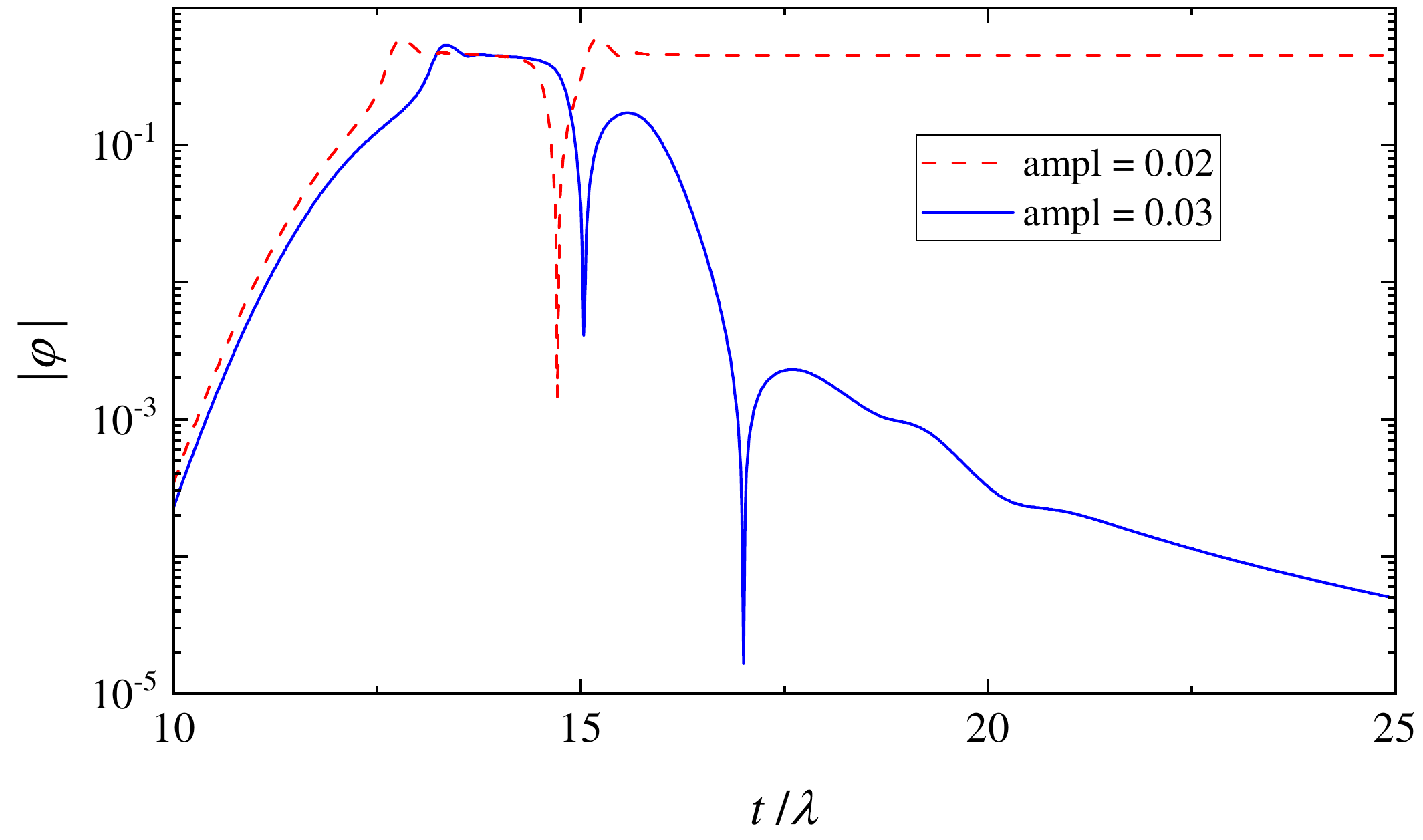}
	\caption{The time evolution of the scalar field on the background of a Schwarzschild black hole with mass $M/\lambda=0.03$ and $\kappa=500$. The initial data is a Gauss pulse with amplitudes 0.03 and 0.02, dispersion $\sigma/\lambda=1$, located at coordinate distance $x/\lambda=12$. The used coupling function is $f_2(\varphi)$. }
	\label{fig:phi_t_phi6_b500}
\end{figure}

It is clear that the condition for the existence of scalarized phases discussed in the main text of the paper can we satisfied also for higher powers of $\varphi$. In the present paper we are focusing on $Z_2$ symmetric theories and that is why the next possible choice is to consider coupling function of the type $f_2(\varphi)$  defined as
\begin{equation}
	f_2(\varphi)= \frac{1}{6\kappa}\left(1- e^{-\kappa\varphi^6}\right).
\end{equation}

The  time evolution of scalar field on the Schwarzschild background is shown in Fig. \ref{fig:phi_t_phi6_b500} for two different amplitudes of the initial perturbation close to the threshold for the development of a nonlinear instability. As one can see, for  smaller amplitude the scalar field perturbation decays exponentially and the quasinormal modes of the Schwarzschild black hole within sGB gravity are observed. A slight increase of the amplitude, though, leads to the formation of a stable equilibrium scalar field configuration at later times.

\subsubsection{Scalarized phases with $f_2(\varphi)$}

\begin{figure}
	\includegraphics[width=0.45\textwidth]{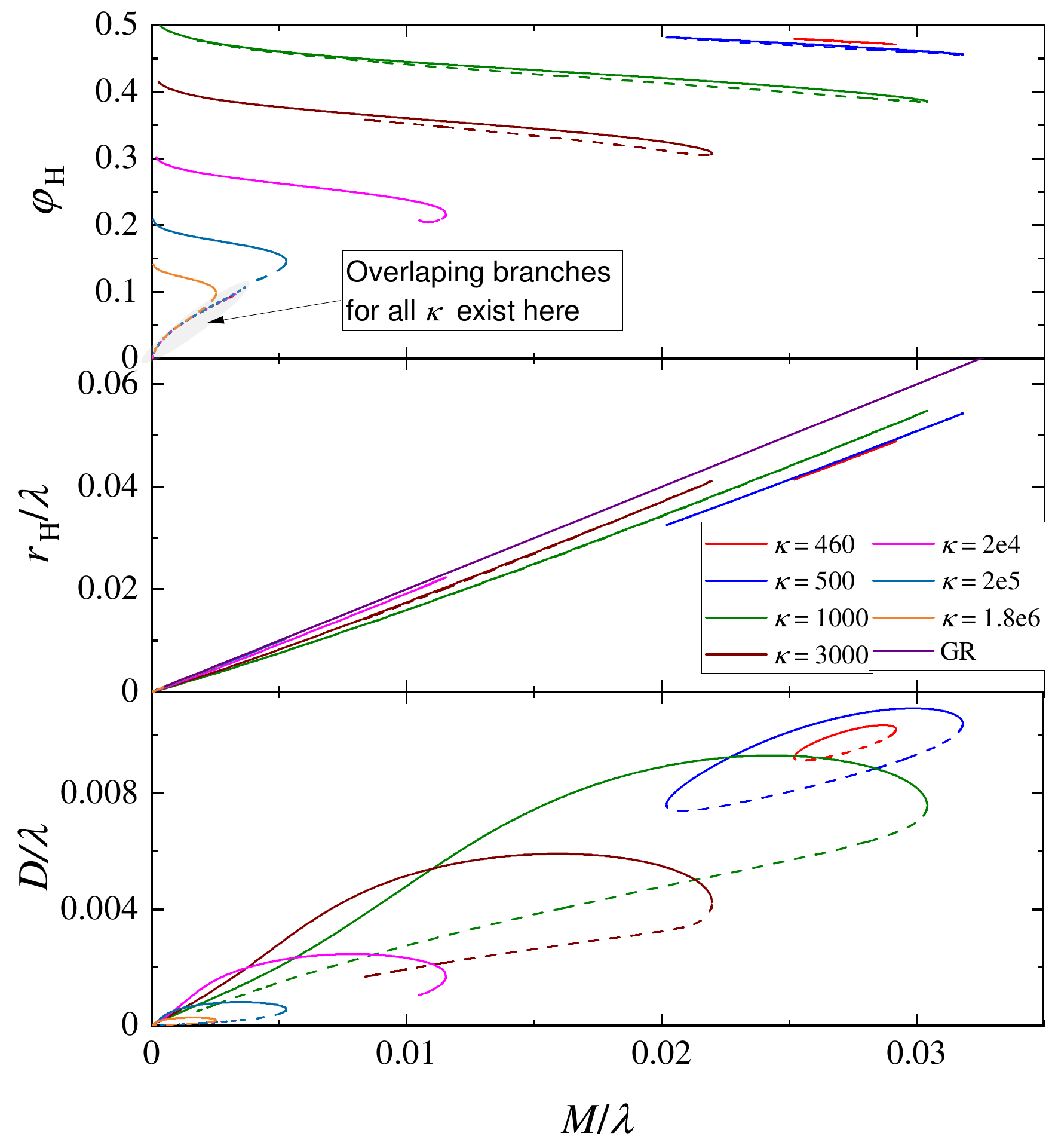}
	\caption{The scalar field at the horizon (top panel), the mass (middle panel) and the scalar charge (bottom panel) as functions of the black hole mass for the $f_2(\varphi)$ coupling. }
	\label{fig:M_phiH_D_phi6}
\end{figure}

Having demonstrated that  scalarized phases can be indeed dynamically formed for large enough amplitude of the initial perturbation, here we will discuss the complete spectrum of scalarized phase solutions for the coupling $f_2(\phi)$  obtained after solving the reduced field equations \eqref{DRFE1}--\eqref{DRFE4}. The scalar field on the horizon, the horizon radius and the scalar charge as a function of mass are plotted in Fig. \ref{fig:M_phiH_D_phi6} for several values of $\kappa$. The minimum $\kappa=460$ plotted in the figure is very close to the threshold where the scalarized phases disappear. Similar to the $f_1(\varphi)$ coupling discussed in the main text, with the increase of $\kappa$ we have three distinct cases 
\begin{itemize}
	\item For smaller $\kappa$ three branches of solutions exist -- two with higher mass spanning a limited range of masses (and not reaching the $M=0$ limit), and one branch appearing at relatively low masses. Only the high-mass branch depicted with solid line is potentially stable.
	\item For intermediate $\kappa$ a similar picture is observed with the difference that the potentially stable branch (depicted with solid line) actually reaches the $M=0$ limit.
	\item For large enough $\kappa$ only two branches of solutions exist starting from the zero mass limit and merging again at some  finite $M/\lambda$. 
\end{itemize}

The main difference with the results for $f_1(\varphi)$ is on quantitative level. Scalarized phases appear for much smaller masses (normalized with respect to $\lambda$) and larger values of $\kappa$. The maximum differences with Schwarzschild is smaller compared to the $f_1(\varphi)$ coupling judging from the middle panel in Fig.~\ref{fig:M_phiH_D_phi6}. Again, even though the differences between some of the branches (for a fixed $\kappa$) look very small if one focuses for example on the behavior of the scalar field at the horizon or the black hole horizon radius, a clear distinction between the scalarized phases is observed in terms of the scalar charge.

The nonlinear simulations show that the only stable branch is the one depicted with a solid line in Fig. \ref{fig:M_phiH_D_phi6}. In addition,  this is the scalarized phase with the highest entropy that is another argument in favor of its stability. Similar to the previous section, for large values of $\kappa$ a small part of the stable hairy black hole branch (close to the maximum mass) can have entropy smaller than the Schwarzschild one but this is quickly reversed with the decrease of the black hole mass. For smaller  $\kappa$  the scalarized phase has always larger entropy with respect to Schwarzschild.

\end{document}